\documentclass[twocolumn,showpacs,preprintnumbers,superscriptaddress,amsmath,amssymb]{revtex4}

\usepackage{graphicx}
\usepackage{dcolumn}
\usepackage{bm}

\begin{document}

\title{High Dimensional Apollonian Networks}

\author{Zhongzhi Zhang}
\affiliation{
Institute of System Engineering, Dalian University of Technology,
2 Ling Gong Rd., Dalian 116023, Liaoning, China}%
\email{xinjizzz@sina.com}

\author{Francesc Comellas}
\affiliation{
Departament de Matem\`atica Aplicada IV, EPSC, Universitat
Polit\`ecnica de Catalunya\\
 Avinguda del Canal Ol\'{\i}mpic s/n, 08860
Castelldefels, Barcelona, Catalonia, Spain
}%
\homepage{http://www-mat.upc.es/~comellas}
\email{comellas@mat.upc.es}

\author{Guillaume Fertin}
\affiliation{
LINA, Universit\'e de Nantes, 2 rue de la Houssini\`ere,
 BP 92208, 44322 Nantes Cedex 3, France
}%
\homepage{http://www.sciences.univ-nantes.fr/info/perso/permanents/fertin/}
\email{fertin@lina.univ-nantes.fr}

\author{Lili Rong}
\affiliation{
Institute of System Engineering, Dalian University of Technology,
2 Ling Gong Rd., Dalian 116023, Liaoning, China}%
\homepage{http://management.dlut.edu.cn:8080/glxy/mydocument/faculty/ronglili.htm}
\email{llrong@dlut.edu.cn}
\date{}

\begin{abstract}
We propose a simple algorithm which produces high
dimensional Apollonian networks with both small-world and
scale-free characteristics. We derive analytical expressions
for the  degree distribution, the clustering
coefficient and the diameter of the networks, which are determined by
their dimension.
\end{abstract}



\maketitle


\section{Introduction}

Since the ground-breaking papers by Watts and Strogatz on
small-world networks \cite{WaSt98} and Barab\'asi and Albert on
scale-free networks \cite{BaAl99}, the research interest on
complex networks as an interdisciplinary subject has soared
\cite{AlBa02,DoMe02,Ne03}. Complex networks describe many systems
in nature and society, most of which share three apparent
features: power-law degree distribution, small average path length
(APL) and high clustering coefficient.

While many models \cite{AlBa02,DoMe02,Ne03} have been proposed to
describe real-life networks, most of them are stochastic. However,
new deterministic models with fixed degree distributions constructed
by recursive methods have been recently introduced
\cite{BaRaVi01,IgYa01,DoGoMe02,JuKiKa02,CoFeRa04,RaBa03,No03}.
Deterministic models have the strong advantage that it is often
possible to compute analytically their properties, which may be
compared with experimental data from real and simulated networks.
Deterministic networks can be created by various techniques:
modification of some regular graphs~\cite{CoOzPe00}, addition and
product of graphs~\cite{CoSa02}, and other mathematical methods as
in~\cite{ZhWaJiHuCh04}. Concerning the problem of Apollonian
packing, two groups independently introduced the Apollonian networks
\cite {AnHeAnSi05, DoMa05} which have interesting properties like
being scale-free, Euclidean, matching, space-filling and can be
applied to porous media, polydisperse packings, road networks or
electrical supply systems \cite {AnHeAnSi05} and may also help to
explain the properties of energy landscapes and the associated
scale-free network of connected minima \cite{DoMa05}.

In this paper we present a simple iterative  algorithm to generate
high-dimensional Apollonian networks based on a similar idea as
that of the recursive graphs proposed in \cite{CoFeRa04}. The
introduced algorithm can concretize the problems of abstract
high-dimensional Apollonian packings. Using the algorithm we
determine relevant characteristics of high-dimensional Apollonian
networks: the degree distribution, clustering coefficient and
diameter, all of which depend on the dimension of Apollonian
packings.

It should be pointed out that the concept of high-dimensional
Apollonian networks was already introduced in \cite{AnHeAnSi05} and
\cite{DoMa05}. In these works, however, the emphasis is placed on
two-dimensional Apollonian networks and their aim is to address the
behavior of dynamical processes \cite{AnHeAnSi05} or provide a model
to help understand the energy landscape networks
\cite{DoMa05,Do02,DoMa05b}. Here, we focus on the producing
algorithm, based on which we provide a detailed calculation of the
topology characterization of high-dimensional Apollonian networks
and we show that it depends on the dimension.

\begin{figure}
\begin{center}
\includegraphics[width=0.3\textwidth]{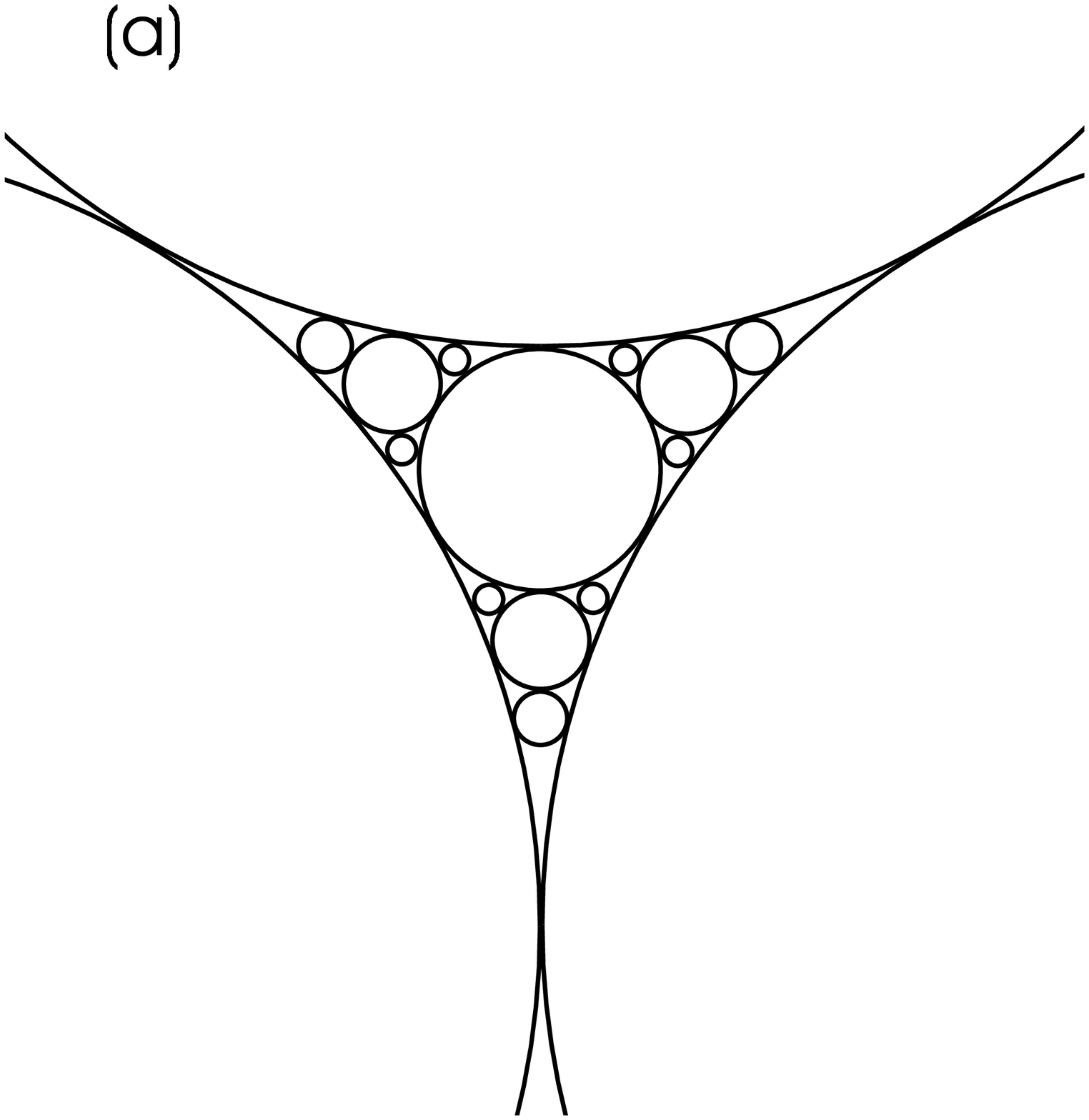} \
\includegraphics[width=0.5\textwidth]{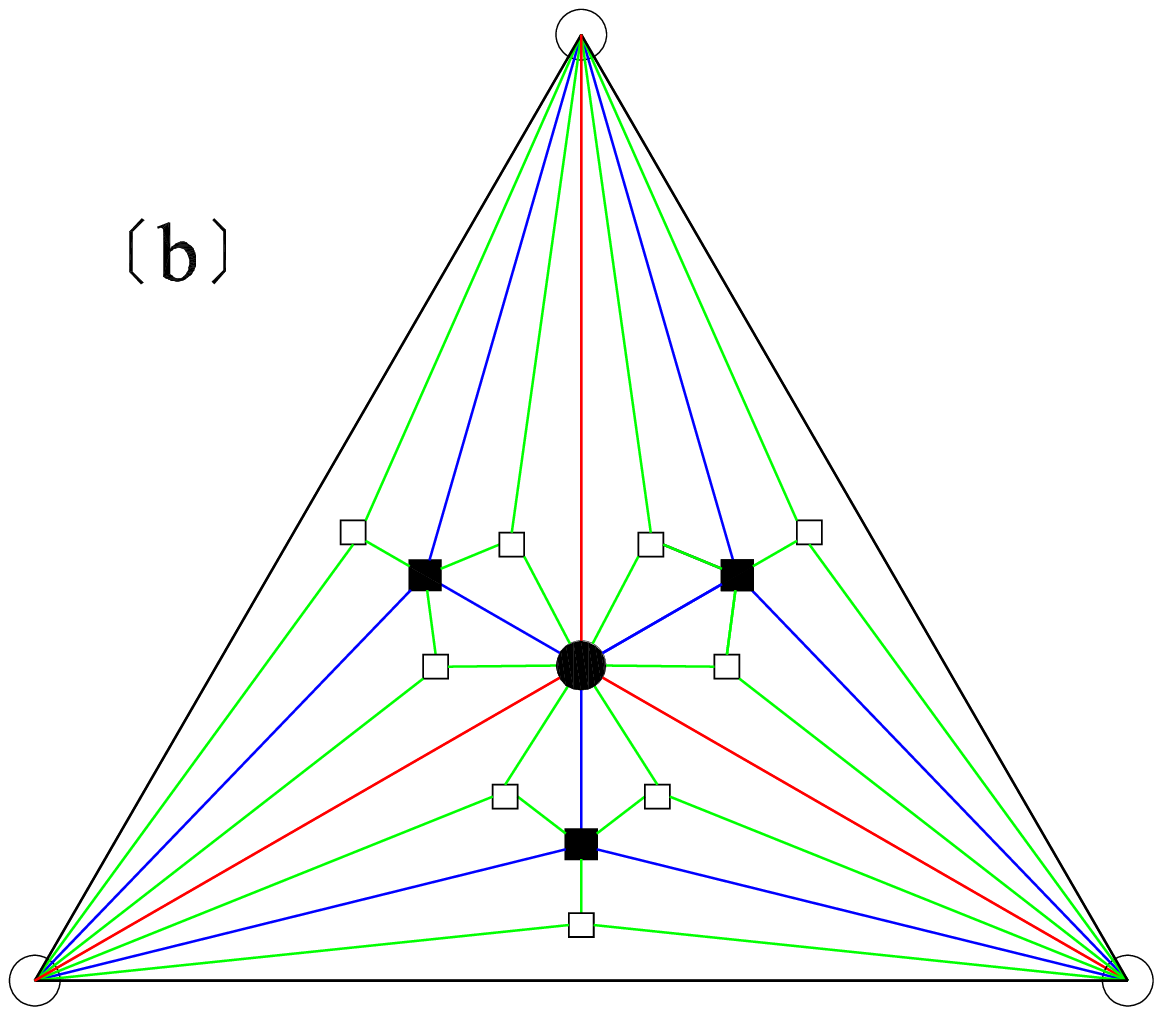}
\end{center}
\caption[kurzform]{\label{Fig1} (a) A two-dimensional Apollonian
packing of disks. (b) Construction of two-dimensional Apollonian
networks, showing the first iterations steps. Vertices introduced in
the initial, first, second and third iterations have symbols
$\bigcirc$, $\bullet$, $\blacksquare$ and $\Box$ respectively.}
\end{figure}

\section{The construction of high dimensional Apollonian networks}
From the problem of Apollonian packing, a two-dimensional example of
which is shown in Fig.1(a), Andrade et al. introduced Apollonian
networks \cite{AnHeAnSi05} which were independently proposed by Doye
and Massen in \cite{DoMa05}. Apollonian packing dates back to
Apollonius of Perga who lived around 200 BC. The classic
two-dimensional Apollonian packing is constructed by starting with
three mutually touching circles, whose interstice is a curvilinear
triangle to be filled. Then a circle is inscribed, touching all the
sides of this curvilinear triangle. We call this the first iteration
$t=1$ and the initial configuration is denoted by $t=0$. For
subsequent iterations we indefinitely repeat the process for all the
newly generated curvilinear triangles. In the limit of infinite
iterations, the well-known two-dimensional Apollonian packing is
obtained.

From the two-dimensional Apollonian packing, one can
straightforwardly define a two-dimensional Apollonian network \cite
{AnHeAnSi05, DoMa05}, where the vertices are associated to the
circles and two vertices are connected if the corresponding circles
are tangent. Figure 1(b) shows how the network is translated from
the two-dimensional Apollonian packing construction. The two-dimensional
Apollonian network can be generalized to high-dimensions
($d$-dimensional, $d\geq 2$) \cite{DoMa05} associated with other
self-similar packings \cite{MaHeRi04}). Next a comprehensive account
will be given.

First, we address $d$-dimensional Apollonian packings which are
constructed iteratively in a similar way as shown in Fig. 1(a).
Initially, we have $d+1$ mutually touching $d$-dimensional
hyperspheres with a curvilinear $d$-dimensional simplex
($d$-simplex) as their interstice. In the first iteration one
$d$-hypersphere is added to fill the interstice in the initial
configuration, such that the added $d$-hypersphere touches each of
the $d+1$ $d$-hyperspheres. The process is continued for all the
newly created curvilinear $d$-dimensional simplex in the successive
iterations. In the limit of infinite iterations, the resulting
pictures are $d$-dimensional Apollonian packings. If each
$d$-hypersphere corresponds to a vertex and vertices are connected
by an edge if the corresponding $d$-hyperspheres are in contact,
then one gets $d$-dimensional Apollonian networks.

\section{The iterative algorithm of high dimensional Apollonian networks}
In the iterative process for the construction of high-dimensional
Apollonian networks at each iteration, for each new hypersphere
added, $d+1$ new interstices are created in the associated
Apollonian packing which will be filled in the next iteration. when
building networks, we can say in equivalent words that for each new
vertex added, $d+1$ new $d$-simplices are generated in the network,
into which vertices will be inserted in the next iteration.
According to this process, we can introduce a general iterative
algorithm to create high-dimensional Apollonian networks which is
similar to the process that allow the construction of the recursive
graphs introduced in \cite{CoFeRa04}.

Before introducing the algorithm we give the following definitions.
A \textbf{complete graph} $K_d$ (also referred in the literature as
\textbf{$d$-clique}; see \cite{We01}) is a graph with $d$ vertices,
where there is no loop or multiple edge and every vertex is joined
to every other by an edge. Generally speaking, two graphs are said
to be \textbf{isomorphic} if the vertices and edges of one graph
match up with vertices and edges of other, and the edge matching be
consistent with the vertex matching.

We denote the $d$-dimensional Apollonian network after $t$
iterations by $A(d,t)$, $d\geq 2, t\geq 0$. Then the $d$-dimensional
Apollonian network at step $t$ is  constructed as follows: For
$t=0$, $A(d,0)$ is the complete graph $K_{d+1}$ (or $(d+1)$-clique),
and $A(d,0)$ has $d+1$ vertices and $(d+1)d\over 2$ edges. For
$t\geq 1$, $A(d,t)$ is obtained from  $A(d,t-1)$ by adding for each
of its existing subgraphs isomorphic to a $(d+1)$-clique and created
at step $t-1$ a new vertex and joining it to all the vertices of
this subgraph (see Fig. 1(b) for the case $d=2$). Then, at $t=1$, we
add one new vertex and $d+1$ new edges to the graph, creating $d+1$
new $K_{d+1}$  cliques and resulting in the complete graph with
$d+2$ vertices, denoted $K_{d+2}$. At $t=2$ we add $d+1$ new
vertices, each of them connected to all the vertices of one of the
$d+1$ cliques $K_{d+1}$ created at $t=1$ introducing  $(d+1)^2$ new
edges, and so on.

Note that the addition of each new vertex leads to $d+1$ new
$(d+1)$-cliques and $d+1$ new edges. So the number of vertices and
edges newly born at step $t_i$ become $L_v(t_i)=(1+d)^{t_i-1}$ and
$L_e(t_i)=(1+d)^{t_i}$, respectively. Therefore, similarly to many
real-life networks such as the World Wide Web, the $d$-dimensional
Apollonian network is a growing network, whose number of vertices
increases exponentially with time.

Thus we can easily see that at step $t$, the Apollonian network
$A(d,t)$  has

\begin{equation}\label{Nt}
N_t=(d+1)+\sum_{t_i=1}^{t}L_v(t_i)=\frac {(d+1 )^{t}-1}{d}+d+1
\end{equation}

vertices and

\begin{equation}\label{Et}
|E|_t =\frac{d(d+1)}{2} + \sum_{t_i=1}^{t}L_e(t_i)=
\frac{d(d+1)}{2}+\frac{(d+1)^{t+1}-d-1}{d}
\end{equation}
edges

The average degree $\overline{k}_t$ is then
\begin{equation}
\overline{k}_t = \frac{2|E|_t}{N_t} =
 \frac {2 (d+1)^{t+1}+d^3+d^2-2d-2}
{ \left( d+1 \right) ^{t}+{d}^{2}+d-1}
\end{equation}

For large $t$ it is approximately $2(d+1)$. We can see when $t$ is
large enough the resulting networks are sparse graphs as many
real-world networks whose vertices have many fewer connections than
is possible.

%
%

\section{Relevant characteristics of high dimensional Apollonian networks}

Below we will find that the dimension $d$ is a tunable parameter
controlling all the relevant characteristics of the $d$-dimensional
Apollonian network.

\subsection{Degree distribution}

When a new vertex $i$ is added to the graph at step $t_i$, it has
degree $d+1$ and forms $d+1$ new $(d+1)$-cliques. From the iterative
algorithm, we can see that each new neighbor of $i$ generated $d$
new $(d+1)$-cliques with $i$ as one vertex of them. In the next
iteration, these $(d+1)$-cliques will introduce new vertices that
are connected to the vertex $i$. Let $k_i(t)$ be the degree of $i$
at step $t$ and $\Delta k_i(t)$ be the increment of $k_i(t)$ minus
$k_i(t-1)$. Then $\Delta k_i(t)$ of vertex $i$ evolves as

\begin{equation}
\Delta k_i(t)=k_i(t)-k_i(t-1)=d \Delta k_i(t-1)
\end{equation}
combining the initial condition  $k_i(t_i)=d+1$ and $\Delta
k_i(t_i)=d+1$, we obtain
\begin{equation}
\Delta k_i(t)=(d+1)d^{t-t_i-1}
\end{equation}
and the degree of vertex $i$ becomes
\begin{equation}
k_i(t)=d+1+\sum_{t_m=t_i+1}^{t}{\Delta
k_i(t_m)=}(d+1)\left(\frac{d^{t-t_i}-1}{d-1}+1\right)
\end{equation}

The distribution of all vertices and their degrees at step $t$ is
given in Table~\ref{tab:table2}, from which we can see that the
degree spectrum of the graph is discrete and some values of the
degree are absent. To relate the exponent of this discrete degree
distribution to the standard $\gamma$ exponent as defined for
continuous degree distribution, we use a cumulative distribution
$P_{cum}(k) \equiv \sum_{k^\prime \geq k} N(k^\prime,t)/N_t \sim
k^{1-\gamma}$. Here $k$ and $k^\prime$ are points of the discrete
degree spectrum. The analytic computation details are given as
follows.

\begin{table}
\caption{\label{tab:table2} Distribution of vertices and their
degrees for  $A(d,t)$  at step $t$.}
\begin{ruledtabular}
\begin{tabular}{lll}
Num. vert.  &  Degree \\
\hline

$d+1$ & $\sum_{j=0}^{t-1}d^{j}+d$ \\
$1$   & $(d+1)(\sum_{j=0}^{t-2}d^{j}+1)$ \\
$d+1$ & $(d+1)(\sum_{j=0}^{t-3}d^{j}+1)$ \\
$(d+1)^{2}$ & $(d+1)(\sum_{j=0}^{t-4}d^{j}+1)$ \\
$\cdots$  &  $\cdots$  &   \\
$(d+1)^{t-3}$ & $(d+1)(d+1+1)$ \\
$(d+1)^{t-2}$ & $(d+1)(d+1)$ \\
$(d+1)^{t-1}$ & $d+1$ \\

\end{tabular}
\end{ruledtabular}
\end{table}


For a degree $k$
\begin{equation*}
k=(d+1)\left(\frac{d^{t-l}-1}{d-1}+1\right)
\end{equation*}
there are  $(d+1)^{l-1}$ vertices with this exact degree, all of
which were introduced at step $l$.

All vertices introduced at time $l$ or earlier have this and a
higher degree. So we have
\begin{equation*}
\sum_{k' \geq k} N(k',t)=(d+1)+\sum_{s=1}^{l}L_v(s)=
\frac{(d+1)^{l}-1}{d}+d+1
\end{equation*}
As the total number of vertices at step $t$ is given
in Eq.~(\ref{Nt}) we have
\begin{eqnarray}
\left[(d+1)(\frac{d^{t-l}-1}{d-1}+1)\right]^{1-\gamma} =
\frac{\frac{(d+1)^{l}-1}{d}+d+1}{\frac{(d+1)^t-1}{d}+d+1}\nonumber\\
=\frac{(d+1)^{l}+d(d+1)-1}{(d+1)^t+d(d+1)-1}
\end{eqnarray}

Therefore, for large $t$ we obtain
\begin{equation*}
(d^{t-l})^{1-\gamma}=(d+1)^{l-t}
\end{equation*}
and
\begin{equation}
\gamma \approx 1+\frac{\ln (d+1)}{\ln d}
\end{equation}
so that $2<\gamma <2.584 96$.
We note that this result was already obtained by Doye and Massen in \cite{DoMa05}.

Also, notice that when $t$ gets large, the maximal degree of a
vertex roughly equals to $d^{t-1} \sim N_t^{\ln d/\ln (d+1)} =
N_t^{1/(\gamma-1)}$.
%


\subsection{Clustering distribution}
The clustering coefficient~\cite{WaSt98} of a single vertex is the
ratio of the total number of edges that actually exist between all
$k$ its nearest neighbors and the potential number of edges
$k(k-1)/2$ between them. The clustering coefficient of the whole
network is the average of all individual vertices. We can derive
analytical expressions for the clustering $C(k)$ for any vertex with
degree $k$.

When a vertex is created it is connected to all the vertices of a
$(d+1)$-clique whose vertices are completely interconnected. It
follows that a vertex with degree $k=d+1$ has a clustering
coefficient of one because all the $(d+1)d/2$ possible links between
its neighbors actually exist. In the following steps, if a newly
created vertex is connected to the considered vertex whose $d$
nearest neighbors must also be linked by the new one at the same
step. Thus for a vertex $v$ of degree $k$, the exact expression for
its clustering coefficient is

\begin{equation}
C(k)= {{{d(d+1)\over 2}+ d(k-d-1)} \over {k(k-1)\over 2}}=
\frac{2d(k-\frac{d+1}{2})}{k(k-1)}
\end{equation}
depending on degree $k$ and dimension $d$. Using this result, we can
compute now the clustering of the graph at step $t$, it is
$\overline{C}_t=S_t/N_t$, where

\begin{equation}
\small S_t =(d+1)\frac{2d(D_0-\frac{d+1}{2})}{D_0(D_0-1)}
     +  \sum_{q=1}^{t} \frac{2d(D_q-\frac{d+1}{2})L_v(q)}{D_q(D_q-1)}
\end{equation}
where $D_0=\frac{d^t-1}{d-1}+d$ and
$D_q=(d+1)\left(\frac{d^{t-q}-1}{d-1}+1\right)$ given by eq. (6) are
the degrees of the vertices created at steps $0$ and $q$,
respectively. One can easily prove that for $t\geq 7 $ and for any
$d\geq 2$ the following relation hold true.

\begin{equation}
\overline{C}_t>\frac{3d-2}{3d-1}
\end{equation}

Therefore the clustering coefficient of high dimension Apollonian
networks is very large. Similarly to the power exponent $\gamma$ of
the degree distribution, the clustering is also tunable simply by
changing the value of control parameter $d$. From Eq. (11), one can
see that the clustering coefficient increases with $d$ and
approaches a limit of $1$ when $d$ gets large. In the special cases
$d=2$ and $d=3$, $\overline{C}$ equals to constant asymptotic values
0.8284~\cite{AnHeAnSi05} and 0.8852, respectively.

\subsection{Diameter}
The diameter of a network characterizes the maximum communication
delay in the network and is defined as the longest shortest path
between all pairs of vertices. In what follows, the notations $
\lceil x \rceil$ and $\lfloor x \rfloor$ express the integers
obtained by rounding $x$ to the nearest integers towards infinity
and minus infinity, respectively. Now we compute the diameter of
$A(d,t)$, denoted $diam(A(d,t))$ for $d\geq 2$ :


{\em Step 1}.  The diameter is $1$.


{\em Steps 2 to $\lceil\frac{d}{2}\rceil+1$}.  In this case, the
diameter is 2, since any new vertex is by construction connected to
a $(d+1)$-clique, and since any $(d+2)$-clique during those steps
contains at least the vertex created at step 1, which is from the
initial clique $K_{d+2}$ or $A(d,1)$ obtained after step 1, thus the
diameter is 2.



{\em Steps $\lceil\frac{d}{2}\rceil+2$ to $d+2$}. In any of those
steps, some newly added vertices might not share a neighbor in the
original clique $K_{d+2}$ obtained after step 1; however, any
newly added vertex is connected to at least one vertex of the
initial clique $K_{d+2}$. Thus, the diameter is equal to 3.


{\em Further steps}. Clearly, at each step $t\geq d+3$, the diameter
always lies between a pair of vertices that have just been created
at this step. We will call the newly created vertices ``outer''
vertices. At any step $t\geq d+3$, we note that an outer vertex
cannot be connected with two or more vertices that were created
during the same step $0<t'\leq t-1$. Moreover, by construction no
two vertices that were created during a given step are neighbors,
thus they cannot be part of the same $(d+2)$-clique. Thus, for any
step $t\geq d+3$, some outer vertices are connected with vertices
that appeared at pairwise different steps. Thus, there exists an
outer vertex $v_t$ created at step $t$, which is connected to
vertices $v_i$s, $1\leq i\leq t-1$, where all the $i$s are pairwise
distinct. We conclude that $v_t$ is necessarily connected to a
vertex that was created at a step $t_0\le t-d-1$. If we repeat this
argument, then we obtain an upper bound on the distance from $v_t$
to the initial clique $K_{d+2}$. Let $t=\alpha (d+1)+p$, where
$2\leq p\leq d+2$. Then, we see that $v_t$ is at distance at most
$\alpha +1$ from a vertex in $K_{d+2}$. Hence any two vertices $v_t$
and $w_t$ in $A(d,t)$ lie at distance at most $2(\alpha +1)+1$;
however, depending on $p$, this distance can be reduced by 1, since
when $p\leq \lceil\frac{d}{2}\rceil+1$, we know that two vertices
created at step $p$ share at least a neighbor in $K_{d+2}$. Thus,
when $2\leq p\leq \lceil\frac{d}{2}\rceil +1$, $diam(A(d,t))\leq
2(\alpha +1)$, while when $\lceil\frac{d}{2}\rceil +2\leq p\leq
d+2$, $diam(A(d,t))\leq 2(\alpha +1)+1$.
One can see that these distance bounds can be reached by pairs of
outer vertices created at step $t$. More precisely, those two
vertices $v_t$ and $w_t$ share the property that they are
connected to $d$ vertices that appeared respectively at steps
$t-1,t-2,\ldots t-d-1$.

Hence, the diameter increases by 2 every $d+1$ steps. More
precisely, we have the following result, for any $d\geq 2$ and
$t\geq 2$ (when $t=1$, the diameter is clearly equal to 1):
$$diam(A(d,t))=2(\lfloor\frac{t-2}{d+1}\rfloor +1)+f(d,t)$$
where $f(d,t)=0$ if $t-\lfloor\frac{t-2}{d+1}\rfloor (d+1)\leq
\lceil\frac{d}{2}\rceil+1$, and 1 otherwise.

In the limit of large $t$, $diam(A(d,t))\sim \frac{2t}{d+1}$, while
$N_t\sim (d+1)^{t-1}$, thus the diameter is small and scales
logarithmically with the network size.

\section{Conclusion and discussion}
In conclusion, we have introduced a general iterative algorithm to
produce high dimensional Apollonian networks associated with high
dimensional packings. The networks present the typical
characteristics of real-life networks in nature and society as they
are small-world  and have a power-law degree distribution. We
compute analytical expressions for the degree distribution, the
clustering coefficient, and the diameter of the networks, all of
which are determined by the dimension of the associated Apollonian
packings. The high dimensional Apollonian networks introduced here,
and the consideration of the method presented in Ref.
\cite{ZhYaWa05}, allow the construction of  high dimensional random
Apollonian networks~\cite{ZhRoCo05}. In addition, it should be worth
studying in detail physical models such as Ising
models~\cite{AnHe05} and processes such as percolation, spreading,
searching and diffusion that take place on the higher-dimensional
Apollonian networks to know also their relation with the dimension.
\smallskip

This research was supported by the National Natural Science
Foundation of China under Grant No. 70431001. Support for F.C. was
provided by the Secretaria de Estado de Universidades e
Investigaci\'on (Ministerio de Educaci\'on y Ciencia),  Spain, and
the European Regional Development Fund (ERDF)  under project
TIC2002-00155. 

\end{document}